\documentclass[doublecol,linenumbers]{epl2} 

\title{Measurements of Newton's gravitational constant and the length of day}
\shorttitle{Measurements of Newton's gravitational constant and the length of day} 

\author{J. D. Anderson\inst{1\footnote{Retired.}} \and G. Schubert\inst{2} \and V. Trimble\inst{3} \and M. R. Feldman\inst{4}}
\shortauthor{J. D. Anderson \etal}

\institute{                    
  \inst{1} Jet Propulsion Laboratory, California Institute of Technology - Pasadena, CA 91109, USA \\
  \inst{2} Department of Earth, Planetary and Space Sciences, University of California, Los Angeles \\ Los Angeles, CA 90095, USA \\
  \inst{3} Department of Physics and Astronomy, University of California Irvine - Irvine CA 92697, USA \\
  \inst{4} Private researcher - Los Angeles, CA 90046, USA
}

\abstract{About a dozen measurements of Newton's gravitational constant, $G$, since 1962 have yielded values that differ by far more than their reported random plus systematic errors. We find that these values for $G$ are oscillatory in nature, with a period of $P = 5.899 \pm 0.062$ yr, an amplitude of $(1.619 \pm 0.103) \times 10^{-14}$~m$^3$~kg$^{-1}$~s$^{-2}$, and mean-value crossings in 1994 and 1997. However, we do not suggest that $G$ is actually varying by this much, this quickly, but instead that something in the measurement process varies. Of other recently reported results, to the best of our knowledge, the only measurement with the same period and phase is the Length of Day (LOD - defined as a frequency measurement such that a positive increase in LOD values means slower Earth rotation rates and therefore longer days). The aforementioned period is also about half of a solar activity cycle, but the correlation is far less convincing. The 5.9 year periodic signal in LOD has previously been interpreted as due to fluid core motions and inner-core coupling. We report the $G$/LOD correlation, whose statistical significance is $0.99764$ assuming no difference in phase, without claiming to have any satisfactory explanation for it. Least unlikely, perhaps, are currents in the Earth's fluid core that change both its moment of inertia (affecting LOD) and the circumstances in which the Earth-based experiments measure $G$. In this case, there might be correlations with terrestrial magnetic field measurements.
}

\begin{document}

\maketitle

\section{Introduction}
Newton's gravitational constant, $G$, is one of a handful of universal constants that comprise our understanding of fundamental physical processes \cite{Mohr2012} and plays an essential role in our understanding of gravitation, whether previously in Newton's attractive gravitational force between two massive bodies $m_1,m_2$ of magnitude \cite{Newton1687}
\begin{equation}
F = \frac{Gm_1 m_2}{r^2},
\end{equation}
where $r$ is their separation distance, or currently as the proportionality constant in the interaction between energy-momentum content $T_{ab}$ (the stress-energy tensor) and space-time curvature $G_{ab}$ (Einstein tensor) in Einstein's general relativity \cite{Einstein1916, Wald1984}
\begin{equation}
G_{ab} = R_{ab} - \frac{1}{2}g_{ab}R= 8\pi G T_{ab},
\end{equation}
in units where the local speed of light in vacuum $c=1$. Yet, experimental determination of Newton's gravitational constant remains a challenging endeavor. As reviewed in \cite{Speake2014}, several measurements over the last thirty years appear to give inconsistent values for $G$, of course an issue for our understanding of this universal constant. Our purpose with this letter is to inform the reader of a one-to-one correlation between an apparent temporal periodicity in measurements of $G$, generally thought to result from inconsistency in measurements, with recently reported oscillatory variations in measurements of LOD \cite{Holme2013}. LOD refers to the excess of the duration of the day (observed period of rotation of the Earth) relative to a standard unit and is calculated by taking the difference between atomic time (TAI) and universal time (UT1) divided by the aforementioned standard unit of $86400$~SI s \cite{Ray1996}. Variations in LOD can be used to determine changes in the Earth's rotation rate effectively providing a means to examine geophysical and atmospheric processes \cite{Peltier2007}.

\begin{figure}
\includegraphics[width=8.0cm]{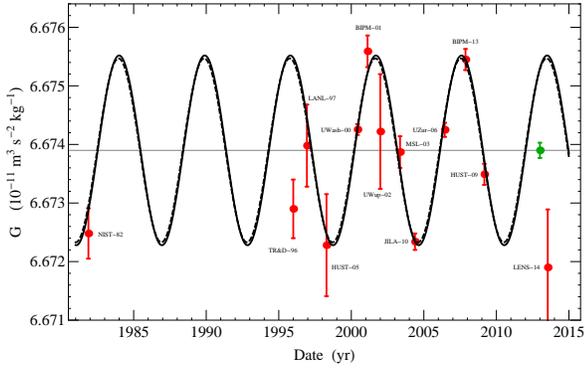}
\caption{Result of the comparison of the CODATA set of $G$ measurements with a fitted sine wave (solid curve) and the 5.9 year oscillation in LOD daily measurements (dashed curve), scaled in amplitude to match the fitted $G$ sine wave. The acronyms for the measurements follow the convention used by CODATA, with the inclusion of a relatively new BIPM result from Quinn {\it et al.}~\cite{Quinn2013} and another measurement LENS-14 from the MAGIA collaboration \cite{Rosi2014} that uses a new technique of laser-cooled atoms and quantum interferometry, rather than the macroscopic masses of all the other experiments. The green filled circle represents the weighted mean of the included measurements, along with its one-sigma error bar, determined by minimizing the L1 norm for all 13 points and taking into account the periodic variation.}
\label{PlotL1}
\end{figure}

For the following discussion, we emphasize that our $G$ analysis and LOD analysis (a verification of the procedures employed in \cite{Holme2013}) are very much independent of one another with the determined fitting parameters for both the period and phase of the periodicities in these measurements coinciding in near perfect agreement. Although we recognize that the one-to-one correlation between the fit to the $G$ measurements and the LOD periodicity of 5.9 years could be fortuitous, we think this is unlikely, given the striking agreement shown in Fig.~\ref{PlotL1}. Furthermore, after taking into account this fitted oscillatory trend in the $G$ measurements, we obtain agreement amongst the different experiments mentioned in \cite{Speake2014} with a weighted mean value for $G$ of $( 6.673899 \pm 0.000069 ) \times 10^{-11}$~m$^3$~kg$^{-1}$~s$^{-2}$.

\section{Methods}

In the July 2014 issue of Physics Today, Speake and Quinn \cite{Speake2014} lay out the problem and review the history of seemingly inconsistent measurements of the gravitational constant $G$. They plot twelve $G$ determinations, along with one-sigma error bars, extending from an experiment by Luther and Towler at the National Bureau of Standards (NBS) in 1982 \cite{Luther1982} to their own at BIPM in 2001 and 2007 (the latter of which was published in 2013) \cite{Quinn2001, Quinn2013}, two measurements in good agreement with each other, but not with the other 10 measurements. Though the vertical scale of years when the measurements were made is not linear, there is a striking appearance of a periodicity running through these values, characterized by a linear drift which suddenly reverses direction and then repeats more than once.

With this pattern in mind, we compute a periodogram for the measured $G$ values versus estimated dates of when the experiments were run. A single clear period of 5.9 years emerges. The data for our $G$ analysis were obtained directly from Table XVII in the 2010 CODATA report published in 2012 \cite{Mohr2012}. There are 11 classical measurements made at the macroscopic level. To those we added two more recent data points, another macroscopic measurement, which we label BIPM-13, and the first ever quantum measurement with cold atoms, labeled LENS-14. Next we used our best estimates of when the experiments were run, not the publication dates, for purposes of generating a measured $G$ value versus date data file, with one-sigma errors included too. These dates were obtained from the respective articles. This gives us the best data set possible, defined by the measured $G$ values used for the CODATA recommendation plus two more published after 2012.

We fit with the raw standard errors, $\sigma_i$, provided with each of the $G$ measurements and used a numerical minimization of the L1 and L2 norms of the weighted residuals, $r_i/\sigma_i$, where the residuals are about a fitting model of a single sine wave, $a_0 + a_1\cos{\omega t}+b_1\sin{\omega t}$, four parameters in all with 13 measurements. Results for the fit to the 13 measured $G$ values are summarized in Fig.~\ref{PlotL1}. The L2 minimization is equivalent to a weighted least squares fit, yet the L1 minimization (solid line in Fig.~\ref{PlotL1}) is a more robust estimator that discriminates against outliers. Both yield excellent fits with a suggestion that two measurements at Moscow \cite{Karagioz1999} and from the MAGIA collaboration \cite{Rosi2014} are outliers. However, the Moscow value is known to suffer from an unexplained temporal drift \cite{Karagioz1999} and the cold-atom value could be fundamentally different ($G$ at the quantum level). Still, we refrain from speculating further on the cold-atom outlier until more microscopic measurements of $G$ are obtained by different experimental groups. The other 11 measurements are consistent with the L1 fitting curve at the one-sigma level or better. Figure~\ref{PlotL1} appears to provide convincing evidence that there exists a 5.9 year periodicity in the macroscopic determinations of $G$ in the laboratory with variations at the level of $\Delta G/G \sim 2.4 \times 10^{-4}$ about a mean value of $6.673899 \times 10^{-11}$~m$^3$~kg$^{-1}$~s$^{-2}$, close to the value recommended by CODATA in 2010 \cite{Mohr2012} but with a much smaller standard error of 10.3 ppm instead of the CODATA recommended error of 120 ppm.

The most accurate determination by the Washington group \cite{Gundlach2000} with a standard error of 14 ppm now falls squarely on the fitting curve. Because the two BIPM measurements were made at the peak of the fitting curve, they now not only agree, but they are consistent with all other measurements. Notably, the measurement with a simple pendulum gravity gradiometer at JILA is no longer  biased to an unacceptably small value, but like the BIPM measurements it falls right on the fitting curve, but at the minimum of the sine wave. The Huazhong measurement is also at the minimum of the curve.

\section{Results}

With the 5.9 year periodicity in the $G$ measurements accepted, the question arises as to what could be the cause and what does it mean. The only thing we can think of is a correlation with a 5.9 year periodicity in the Earth's LOD, published by Holme and de Viron last year \cite{Holme2013}. The International Earth Rotation and Reference Systems Service (IERS), established in 1987, maintains downloadable data files containing daily values of several parameters related to Earth orientation and rotation. The files extend from 1962 January 01, when the Consultative Committee on International Radio (CCIR) established Universal Time Coordinated (UTC) as the standard for time keeping, to the most current date available. We extract two rotation files, the first is the difference UT1-UTC in seconds and the second the LOD, also expressed in seconds, along with daily estimates of standard errors for both. There is also a piecewise constant file in integer seconds for the standard of atomic time TAI minus UTC. By differencing these two files the phase of the Earth rotation is obtained as measured against a uniform atomic time. This difference can be thought of as a continuous phase function $\phi (t)$ in radians sampled once per day at the beginning of the day. It can be expressed in SI seconds, the units on the IERS files, by multiplying by the conversion factor $86400/2 \pi $. It essentially provides the time gained or lost over the years by a poor mechanical clock, the Earth, which runs slow with a loss of about 33 s over the 52 years of the downloaded file. Because of its name and units of seconds only, the second file LOD is more difficult to interpret. It is also the gain or loss of time by the Earth, but only over the current day, and because of definitions there is a reversal in sign. When expressed as a continuous function of the Earth's rotational frequency $\nu ( t )$, it is simply $\nu_0 - \dot{\phi} / 2 \pi$, where $ \nu_0$ is an adopted frequency of rotation with sidereal period of 86164.098903697 s. The quantity $\dot{\phi}/ ( 2 \pi \nu_0 )$ is small and can be taken to the first order in all calculations.

\begin{figure}
\includegraphics[width=8.0cm]{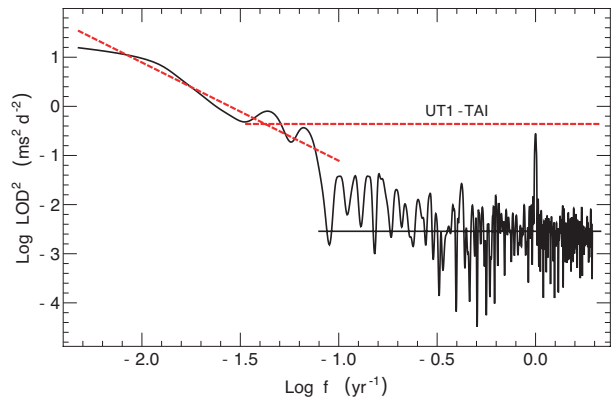}
\caption{One-sided power spectral density per unit frequency for LOD data over the years 1962 to 2014. The white-noise floor is indicated by the horizontal solid line and corresponds to a standard deviation of 0.54~ms~d$^{-1}$, achieved by introduction of GPS data in 1993 and consistent with the daily estimates of standard error archived with the LOD data. The upper dashed curve corresponds to mean spectral density for the numerical time derivative of the UT1 data, dependent on VLBI data from radio sources on the sky. For the low end of the spectrum the LOD and UT1 data both indicate a $f^{-2}$ random walk, which with only 52 years of data can be  confused with a drift in the Earth's rotation.  At the high end, the underlying spectrum indicates white LOD noise, but with a rich spectrum from tidal torques and atmospheric loading at higher frequencies not plotted. Although there is power in the region, there is no suggestion of a single spectral line from the 5.9 year oscillation, a term which must be extracted by analysis in the time domain \cite{Holme2013}.}
\label{PlotLOD}
\end{figure}

Formally, the spectral density of frequency is related to the spectral density of phase by $S_{\mathrm{LOD}} ( f) = ( 2 \pi f )^2 S_{\mathrm{UT1}} ( f )$, where $f$ is the Fourier frequency. However, a separate computation of the spectrum for each file shows that before 1994 either file can be used for analysis, but after the introduction of Global Positioning (GPS) data in 1993, the LOD data become more accurate by a factor of seven or more. This conclusion is consistent with the standard errors included with the data files of LOD and UT1-UTC. We show our estimate of the spectral density for the LOD data in Fig.~\ref{PlotLOD}, obtained by weighted least squares and SVD, but this time with 850 Fourier coefficients, 430 degrees of freedom, and 19169 observations. The spectral resolution is $0.019$~yr$^{-1}$, which we oversample by a factor of four, and the frequency cut off is 2 yr$^{-1}$, far short of the Nyquist frequency of 0.5 d$^{-1}$. A window function is not applied to the data. It introduces undesirable artifacts into the low-frequency noise spectrum of interest and does little to isolate spectral lines. The Gaussian window produces a hint of a line at 5.9 yr, but only a hint. We proceed to an analysis of the data in the time domain.

The 5.9 year periodicity in the LOD data is plotted by Holme and de Viron in Figure 2 of their paper \cite{Holme2013}. Their plot looks in phase with the fit to the 13 $G$ values, but in order to obtain an independent check on the reality of the signal and for purposes of having a numerical sine wave extending into 2014, we first smooth the LOD data with a Gaussian filter with a radius of 600 days and a standard deviation of 200 days. As a result, the high-frequency noise at a period of one year and shorter is practically eliminated, and with little effect on the low-frequency noise spectrum. Next we fit a cubic spline to the smoothed data with a selection of knots or segments for the cubic polynomials done by eye, such that the fitting curve is sufficiently smooth but with a negligible effect on the 5.9 year periodicity. The resulting LOD residuals are fit with a sine wave of fixed 5.9 year period which is then subtracted from the smoothed data. The same procedure is applied to the new smoothed data and the procedure repeated four times with the knots for the spline at closer spacing with each iteration. The final result is the pure sine wave plotted as a dashed curve in Fig.~\ref{PlotL1}. It agrees with the periodic signal found by Holme and de Viron. A removal of the fitted spline representation of the random walk, and also the sine wave, from the smoothed data is all that is needed in order to reduce the LOD residuals about the fit to a one-sigma noise level of $4.8$~$\mu$s~d$^{-1}$. The amplitude of the fitted periodic signal is $92.64\pm 0.18$~$\mu$s~d$^{-1}$, reduced from the amplitude of $150$~$\mu$s~d$^{-1}$ \cite{Holme2013} by the Gaussian smoothing, but with a well-determined period of $5.90076 \pm 0.00074$~yr. With 99\% confidence the period lies between 5.898 and 5.903~yr. The phasing of the sine wave is as shown in Fig.~\ref{PlotL1} with a standard error of 0.25~yr.

\begin{figure}
\includegraphics[width=8.0cm]{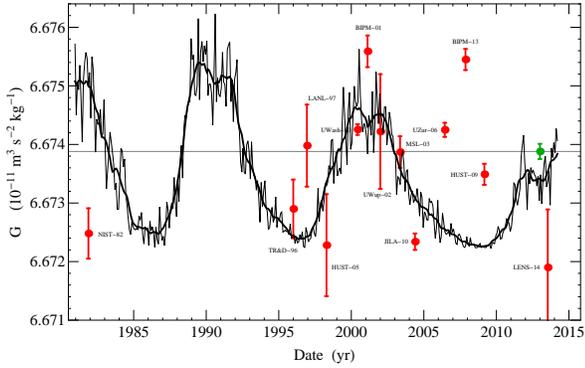}
\caption{Result of the comparison of our $G$ data set with the monthly mean of the total sunspot number, appropriately scaled. The black curves represent solar activity as reflected in the international sunspot number.}
\label{Plotssn}
\end{figure}

The correlation between LOD and $G$ measurements in Fig.~\ref{PlotL1} is most likely of terrestrial origin, but the period of 5.9 years is also close to one-half the principal period of solar activity. References \cite{Djurovic1996} and \cite{Rio2003} discuss in greater detail that a possible correlation between solar activity and LOD measurements is not unexpected. Solar activity has an effect on mass distribution in the atmosphere which ultimately affects the Earth's axial moment of inertia. It is feasible that this effect occurs at longer periods in the 5.9-year range, as well as at much shorter periods, on the order of days, for which models exist \cite{Holme2013}.

Consequently, we plot in Fig.~\ref{Plotssn} the monthly mean of the total sunspot number and also a 13-month smoothing curve, both shown in black. The two curves, again scaled to the magnitude of the $G$ data, are taken directly from freely available downloads of data archived at www.sidc.be by WDC-SILSO, Royal Observatory of Belgium, Brussels. The smoothing is done by a standard tapered-boxcar approach and is generally regarded as a good measure of solar activity. Although the $G$ measurements show a general agreement with solar cycle 23, which peaked around 2002, the long and unexpected minimum that followed, and lasted until about 2010, is at odds with the rise in $G$ values during that minimum. There is also a negative correlation between the measurement from 1982 at the National Bureau of Standards, labeled NIST-82, and the sunspot number. It seems that solar activity can be disregarded as a cause of the variations in $G$ measurements.

\section{Conclusions}

Over the relatively short time span of 34 years considered here, variations in the rotation of the Earth can be considered either a random walk or possibly a drift. Over much longer time scales the rotation must be slowing because of the transfer of spin angular momentum to orbital angular momentum caused by tidal friction of the Moon. Similarly, a real increase in $G$ should pull the Earth into a tighter ball with an increase in angular velocity and a shorter day due to conservation of angular momentum, contrary to the correlation shown in Fig.~\ref{PlotL1}. Thus, we do not expect that this behavior necessarily points to a real variation in $G$ but instead to some yet-to-be determined mechanism affecting both measurements in a similar manner.

Importantly, if the observed effect is connected with a centrifugal force acting on the experimental apparatus, changes in LOD are too small by a factor of about $10^5$ to explain the changes in $G$ for the following reason. The Earth's angular velocity $\omega_E$ is by definition
\begin{equation}
\omega_E = \omega_0 ( 1 - \mathrm{LOD} ),
\end{equation}
where $\omega_0$ is an adopted sidereal frequency equal to $72921151.467064$ picoradians per second and the LOD is in ms~d$^{-1}$ (www.iers.org). The total centrifugal acceleration is given by
\begin{equation}
a_c = r_s \omega_0^2 \bigg[ 1 - 2 A \sin\bigg(\frac{2\pi}{P} (t-t_0)\bigg) \bigg],
\end{equation}
where $A$ is the amplitude $0.000150/86400$ of the 5.9 year sinusoidal LOD variation and $r_s$ is the distance of the apparatus from the Earth's spin axis. The maximum percentage variation of the LOD term is $ 3.47 \times 10^{-9}$ of the steady-state acceleration, while $\Delta G/G$ is $2.4 \times 10^{-4}$, hence even the full effect of the acceleration with no experimental compensation changes $G$ by only $10^{-5}$ of the amplitude in Fig.~\ref{PlotL1}. Perhaps instead, the effect is connected with changing torques on the Earth's mantle due to changing motions in the core. Changes of circulation in the core must be accompanied by changes in density variations in the core causing variations in the gravitational acceleration $g$ in the laboratory. At least this mechanism links both LOD and gravitational changes to changes in the core although we do not immediately see how either of these mechanisms could affect measurements of $G$ in the laboratory given the torsion balance schemes employed.

The least likely explanation is a new-physics effect that could make a difference in the macroscopic and microscopic determinations of $G$. Perhaps a repetition of the single 2014 quantum measurement over the next decade or so can show consistency with a constant value, although if the variations in $G$ measurements are caused by an unknown inertial or frame effect, not by systematic experimental error, it likely applies at both the macroscopic and the microscopic levels. The gravitational parameter for the Sun, $GM_{\odot}$, is known to ten significant figures from orbital motions in the Solar System (ssd.jpl.nasa.gov/?constants). The universal constant $G$ does not vary at that scale, although Krasinsky and Brumberg \cite{Krasinsky2004,Anderson2010} report a detection of an unexplained secular increase in the astronomical unit (AU) over the years 1976 to 2008, which can be interpreted as an increase in $GM_{\odot}$ proportional to the cube of the AU. However the effect on $G$, if real, is at the level of an increase of 3 parts in $10^{12}$ per year and undetectable with laboratory measurements of $G$. Nevertheless, the increase in $GM_{\odot}$ is not explainable as an increase of the solar mass by accretion as opposed to the mass radiated away by solar luminosity \cite{Anderson2010}. Apparently, there does seem to be a secular or very long period (greater than 20000 years) $G$ variation in the Solar System, but of order $10^{-6}$ smaller than the variation shown in Fig.~\ref{PlotL1}.

\acknowledgments

\section{Appendix}
Since the recent publication of this article suggesting that there is a strong correlation with coefficient 0.99764 between measured values of the gravitational constant G as adopted by CODATA \cite{Mohr2012} and the 5.9-year oscillation of the length of day (LOD) \cite{Holme2013}, Schlamminger et al. \cite{Schlamminger2015} point out that the dates assigned to some of the measurements do not agree with the intervals over which the measurements were made. They provide in their TABLE II a corrected and augmented list of measurements carried out over the last 35 years. There are 19 independent measurements at average times accurate to about 20\% of the total interval for each experiment's operation. The BIPM-01 and BIPM-13 values used by CODATA are separated into measurements in the electrostatic servo and Cavendish modes. The HUST-09 data are separated into two segments, one in 2007 with an interval of 60 days and another in 2008 with an interval of 39 days. Three measurement intervals, not included in the CODATA set, from the University of California Irvine (UCI) are appended \cite{Newman2014}. A detailed discussion of the CODATA TR\&D-96 measurement is included for a measurement interval of 3835 days from 1985 to 1995. As this measurement is relatively uncertain (see Fig.~\ref{PlotL1}), and in addition there is little or no evidence of a variation over 10.5 years, we exclude it from the revised data set.

\begin{figure}
\includegraphics[width=8.0cm]{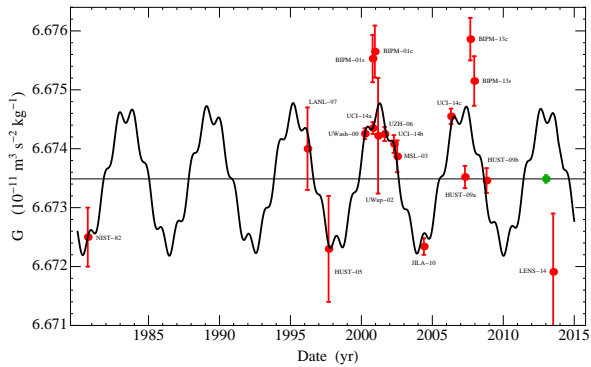}
\caption{Two-period fit to 18 revised $G$ measurements recommended by Schlamminger, Gundlach and Newman \cite{Schlamminger2015}. The L1 norm of the weighted residuals is minimized at a value of 28.81, with the seven parameters of the fitting model and their one-sigma errors from the converged covariance matrix given in Table I. The revised weighted mean and its uncertainty is indicated by the green dot.}
\label{newfit}
\end{figure}

The fit to the remaining 18 points is shown in Fig.~\ref{newfit}, where two periods are included, the original period of about 5.9 years and an annual term suggested by FIG. 3 in Schlamminger et al. \cite{Schlamminger2015}. The amplitude of the periodic fitting curve is reduced significantly, but it has advantages not apparent in the fit to the uncorrected data of Fig.~\ref{PlotL1}, in particular an excellent fit to the most accurate measurements at the maximum between 2000 and 2003, and a reconciliation of the comparably accurate JILA-10 and UCI-14c measurements, which without reconciliation differ by 16 sigma. The LENS-14 measurement is no longer an obvious outlier, but is negatively biased by 2.7 sigma from the fitting curve, no worse than the positive bias in the four BIPM measurements, with only BIPM-13c greater than three sigma. The only obvious outlier is HUST-01a, with a negative six-sigma bias from the curve.

There are seven parameters in the revised fitting model with two periods, as opposed to the simple sine wave of Fig.~\ref{PlotL1} with four parameters. Values from a minimization of the L1 norm and standard errors from the converged covariance matrix are given in Table~1. The fitting model is given by,
\begin{eqnarray}\label{GMdl}
G = a_0 + a_1 \cos\bigg( \frac{2 \pi t}{P_1}\bigg) + b_1 \sin \bigg( \frac{2 \pi t}{P_1} \bigg)  \nonumber
\\ + a_2 \cos \bigg( \frac{2 \pi t}{P_2} \bigg) + b_2 \sin \bigg( \frac{2 \pi t}{P_2} \bigg).
\end{eqnarray}
The two-period model is no longer in phase with the LOD sine wave, and as a consequence the sample correlation coefficient is reduced to 0.860. However, if the LOD sine wave is shifted earlier in phase by 174 days, the correlation coefficient is 0.944. Nevertheless, this makes the interpretation of a possible correlation of the $G$ measurements with LOD more problematic, with the similar periods near 5.9 years possibly a coincidence.

\begin{table}[htbp]
\begin{center}
    \begin{tabular}{ | c | l | c |}
    \hline
    Parameter & Value & Standard Error\\ \hline
   $\rm a_0$ & 6.673488 & 0.000071 \\
   $\rm a_1$ & 0.000084 & 0.000031 \\
   $\rm b_1$ & 0.000150 & 0.000072 \\
   $\rm P_1$ & 1.023087 & 0.000042 \\
   $\rm a_2$ & -0.001116 & 0.000091 \\
   $\rm b_2$ & -0.000126 & 0.000070 \\
   $\rm P_2$ & 5.911615 & 0.000028 \\
    \hline
    \end{tabular}
\end{center}
\caption{Parameters of the fitting model of Eq.~\ref{GMdl}. The periods $\rm P_1$ and $\rm P_2$ are in years and the coefficients are in the units of $G$, or $\rm 10^{-11}~m^3~kg^{-1}~s^{-2}$.}
\end{table}

\end{document}